\documentstyle[epsfig,12pt,a4p]{article}

\newcommand{\pic}{\mbox{$\pi^+\pi^-\pi^+\pi^-$ }}
\newcommand{\pim}{\mbox{$\pi^+\pi^-\pi^0\pi^0$ }}

\begin{document}
\begin{titlepage}
\def\footnoterule{\hrule width 1.0\columnwidth}
\begin{tabbing}
put this on the right hand corner using tabbing so it looks
 and neat and in \= \kill
\> {6 January 2000}
\end{tabbing}
\bigskip
\bigskip
\begin{center}{\Large  {\bf A study of the
$f_0(1370)$, $f_0(1500)$, $f_0(2000)$ and $f_2(1950)$ observed in the
centrally produced
$4\pi$ final states}
}\end{center}

\bigskip
\bigskip
\begin{center}{        The WA102 Collaboration
}\end{center}\bigskip
\begin{center}{
D.\thinspace Barberis$^{  4}$,
F.G.\thinspace Binon$^{   6}$,
F.E.\thinspace Close$^{  3,4}$,
K.M.\thinspace Danielsen$^{ 11}$,
S.V.\thinspace Donskov$^{  5}$,
B.C.\thinspace Earl$^{  3}$,
D.\thinspace Evans$^{  3}$,
B.R.\thinspace French$^{  4}$,
T.\thinspace Hino$^{ 12}$,
S.\thinspace Inaba$^{   8}$,
A.\thinspace Jacholkowski$^{   4}$,
T.\thinspace Jacobsen$^{  11}$,
G.V.\thinspace Khaustov$^{  5}$,
J.B.\thinspace Kinson$^{   3}$,
A.\thinspace Kirk$^{   3}$,
A.A.\thinspace Kondashov$^{  5}$,
A.A.\thinspace Lednev$^{  5}$,
V.\thinspace Lenti$^{  4}$,
I.\thinspace Minashvili$^{   7}$,
J.P.\thinspace Peigneux$^{  1}$,
V.\thinspace Romanovsky$^{   7}$,
N.\thinspace Russakovich$^{   7}$,
A.\thinspace Semenov$^{   7}$,
P.M.\thinspace Shagin$^{  5}$,
H.\thinspace Shimizu$^{ 10}$,
A.V.\thinspace Singovsky$^{ 1,5}$,
A.\thinspace Sobol$^{   5}$,
M.\thinspace Stassinaki$^{   2}$,
J.P.\thinspace Stroot$^{  6}$,
K.\thinspace Takamatsu$^{ 9}$,
T.\thinspace Tsuru$^{   8}$,
O.\thinspace Villalobos Baillie$^{   3}$,
M.F.\thinspace Votruba$^{   3}$,
Y.\thinspace Yasu$^{   8}$.
}\end{center}

\begin{center}{\bf {{\bf Abstract}}}\end{center}

{
The production and decay properties of the
$f_0(1370)$,
$f_0(1500)$, $f_0(2000)$ and $f_2(1950)$ have been
studied in central pp interactions at 450 GeV/c.
The $dP_T$, $\phi$ and $|t|$ distributions of these resonances are
presented.
For the $J$~=~0 states, the $f_0(1370)$ and $f_0(2000)$
have similar $dP_T$ and $\phi$ dependences. These are different
to the $dP_T$ and $\phi$ dependences of the $f_0(980)$, $f_0(1500)$
and $f_0(1710)$. For the $J$~=~2 states the
$f_2(1950)$ has different dependences to the
$f_2(1270)$ and $f_2^\prime(1520)$.
This shows that the $dP_T$ and $\phi$ dependences are not just $J$ phenomena.
}
\bigskip
\bigskip
\bigskip
\bigskip\begin{center}{{Submitted to Physics Letters}}
\end{center}
\bigskip
\bigskip
\begin{tabbing}
aba \=   \kill
$^1$ \> \small
LAPP-IN2P3, Annecy, France. \\
$^2$ \> \small
Athens University, Physics Department, Athens, Greece. \\
$^3$ \> \small
School of Physics and Astronomy, University of Birmingham, Birmingham, U.K. \\
$^4$ \> \small
CERN - European Organization for Nuclear Research, Geneva, Switzerland. \\
$^5$ \> \small
IHEP, Protvino, Russia. \\
$^6$ \> \small
IISN, Belgium. \\
$^7$ \> \small
JINR, Dubna, Russia. \\
$^8$ \> \small
High Energy Accelerator Research Organization (KEK), Tsukuba, Ibaraki 305-0801,
Japan. \\
$^{9}$ \> \small
Faculty of Engineering, Miyazaki University, Miyazaki 889-2192, Japan. \\
$^{10}$ \> \small
RCNP, Osaka University, Ibaraki, Osaka 567-0047, Japan. \\
$^{11}$ \> \small
Oslo University, Oslo, Norway. \\
$^{12}$ \> \small
Faculty of Science, Tohoku University, Aoba-ku, Sendai 980-8577, Japan. \\
\end{tabbing}
\end{titlepage}
\setcounter{page}{2}
\bigskip
\par
The WA102 collaboration  has recently published a study of the
centrally produced $4\pi$ final states~\cite{pi4papc}.
In this paper the production and decay properties of the resonances
observed in these channels will be presented. In previous
publications the properties of the $f_1(1285)$~\cite{f1pap},
$\eta_2(1645)$ and $\eta_2(1870)$~\cite{etapipipap} have already
been presented. In this paper
the properties of the $f_0(1370)$, $f_0(1500)$, $f_0(2000)$ and
$f_2(1950)$ will be discussed.
\par
In previous analyses it has been observed that
when the centrally produced system has been analysed
as a function of the parameter $dP_T$, which is the difference
in the transverse momentum vectors of the two exchange
particles~\cite{WADPT,closeak},
all the undisputed
$ q \overline q $ states
(i.e. $\eta$, $\eta^{\prime}$, $f_1(1285)$ etc.)
are suppressed at small
$dP_T$ relative to large $dP_T$,
whereas the glueball candidates
$f_0(1500)$, $f_0(1710)$ and $f_2(1950)$ are prominent~\cite{memoriam}.
\par
In addition, an interesting effect has been observed in
the azimuthal angle $\phi$ which is defined as the angle between the $p_T$
vectors of the two outgoing protons.
For the resonances
studied to date which are compatible with
being produced by DPE,
the data~\cite{phiangpap} are consistent with the Pomeron
transforming like a non-conserved vector current~\cite{clschul}.
In order to determine the
$\phi$ dependence for the resonances observed,
a spin analysis has been performed on the \pic and \pim channels
in four different $\phi$ intervals each
of 45 degrees.
As an example,
fig.~\ref{fi:1}
shows the $J^{PC}$~=~$0^{++}$ $\rho \rho$ wave from the \pic channel
in the four intervals.
The waves have been fitted in each interval
with the parameters of the resonances fixed to those obtained from the
fits to the total data as described in ref~\cite{pi4papc}.
The distributions found are consistent for the two
channels and
the fraction of each resonance as a function of $\phi$ from the \pic channel
is plotted in
fig.~\ref{fi:2}.
The distributions observed for the $f_0(1370)$ and $f_0(1500)$ are similar
to what was found in the analysis of the $\pi^+ \pi^-$ final
state~\cite{pipikkpap}.
\par
In order to calculate the contribution of each resonance as a function
of $dP_T$, the waves have been fitted
in three $dP_T$ intervals with
the parameters of the resonances fixed to those obtained from the
fits to the total data as described in ref~\cite{pi4papc}.
Table~\ref{ta:dpt} gives the percentage of each resonance
in three $dP_T$ intervals together with the ratio of the number of events
for $dP_T$ $<$ 0.2 GeV to
the number of events
for $dP_T$ $>$ 0.5 GeV for each resonance considered.
The dependences found for the $f_0(1370)$ and $f_0(1500)$ are similar
to what was found in the analysis of the $\pi^+ \pi^-$ final
state~\cite{pipikkpap}.
\par
The fact that the $f_0(1370)$ and $f_0(1500)$ have different
$\phi$ and $dP_T$ dependences confirms that these are not
simply $J$ dependent phenomena. This is also true for the
$J$~=~2 states, where the $f_2(1950)$ has different dependences to the
$f_2(1270)$ and $f_2^\prime(1520)$~\cite{pipikkpap}.
\par
In order to determine the
four momentum transfer dependence ($|t|$) of the
resonances observed
in the \pic channel
the waves have been fitted in 0.1 GeV$^2$ bins
of $|t|$
with the parameters of the resonances fixed to those obtained from the
fits to the total data as described in ref~\cite{pi4papc}.
Fig.~\ref{fi:2} shows the four momentum transfer from
one of the proton vertices for these resonances.
The distributions
have been fitted with a single exponential
of the form $exp(-b |t|)$ and the
values of $b$ found are given in table~\ref{ta:b}.
The values of $b$ for the $f_0(1370)$ and $f_0(1500)$ are similar
to what was found in the analysis of the $\pi^+ \pi^-$ final
state~\cite{pipikkpap}.
\par
The $\phi$ distribution, the $dP_T$ and $t$ dependence of the
$f_2(1950)$ are different to what has
been observed for other $J^{PC}$~=~$2^{++}$
resonances~\cite{pipikkpap} but are similar to what
was observed for the
$\phi \phi$~\cite{phiphipap}
and
$K^*(892) \overline K^*(892)$~\cite{kstkstpap}
final states which were both found to have $J^{PC}$~=~$2^{++}$.
In order to see if the
$\phi \phi$ and
$K^*(892) \overline K^*(892)$
final states could be due to the $f_2(1950)$,
the parameters of the $f_2(1950)$ have been used as input to
a Breit-Wigner function which has been modified to take into account
the different thresholds.
\par
Superimposed on the $\phi \phi$ mass spectrum
in fig.~\ref{fi:3}a) is the distribution that could be due to the
$f_2(1950)$. As can be seen, although the $f_2(1950)$ can describe most of the
spectrum, there is an excess of events in the 2.3 GeV mass region.
Including a Breit-Wigner to describe the $f_2(2340)$,
which has previously been observed decaying to $\phi \phi$~\cite{PDG98},
with
M~=~2330~$\pm$~15~MeV
and $\Gamma$~=~130~$\pm$~20~MeV gives the distribution in
fig.~\ref{fi:3}b).
Assuming that the $f_2(1950)$ has a
$\phi \phi$ decay mode then
correcting for the unseen decay modes the branching ratio of the
$f_2(1950)$ to $f_2(1270)\pi \pi/\phi \phi$ was found to be
72~$\pm$~9.
\par
Superimposed on the $K^{*0} \overline K^{*0}$ mass spectrum
in fig.~\ref{fi:3}c) is the distribution that could be due to the
$f_2(1950)$. As can be seen the $f_2(1950)$ can describe all the
$K^{*0} \overline K^{*0}$ mass spectrum.
Assuming that the $f_2(1950)$ has a
$K^{*0} \overline K^{*0}$ decay mode then
correcting for the unseen decay modes the branching ratio of the
$f_2(1950)$ to $f_2(1270)\pi \pi / K^{*0} \overline K^{*0}$
was found to be
33~$\pm$~4.
In addition,
the branching ratio of the $f_2(1950)$ to $\phi \phi/K^{*0} \overline K^{*0}$
above the $\phi \phi$ threshold is 0.8~$\pm$~0.14.
\par
We have previously published a paper describing the decays of the
$f_0(1370)$ and $f_0(1500)$ to $\pi \pi $ and $K \overline K$~\cite{pipikkpap}.
In ref.~\cite{pi4papc} a fit has been performed to the
$\rho \rho$ and $\sigma \sigma$ final states and the
contributions of the $f_0(1370)$ and $f_0(1500)$ has been determined.
After correcting for the unseen decay modes and the $\sigma \sigma$
decay mode
the branching ratio of the $f_0(1500)$ to $4\pi/\pi\pi$ is found to
be 1.37~$\pm$~0.16 . In the initial Crystal Barrel
publication this value was 3.4~$\pm$~0.8~\cite{cb4pi}.
In the latest
preliminary analysis~\cite{cbhad99}
of the Crystal Barrel data
the value is 1.54~$\pm$~0.6.
Hence although the experiments disagree about
the relative amount of $\rho \rho$ and $\sigma \sigma$ in
the $4\pi$ decay mode~\cite{pi4papc},
the overall measured
branching ratio is consistent.
\par
After correcting for the unseen decay modes and taking into account the
above uncertainties
the branching ratio of the $f_0(1370)$ to $4\pi/\pi\pi$ is found to be
34~$^{+22}_{-9}$.
The large error is due to the fact that
there is considerable uncertainty in the
amount of $f_0(1370)$ in the $\pi\pi$ final state due to the
possible contribution from
the high mass side of the $f_0(1000)$.
In the latest
preliminary analysis~\cite{cbhad99}
of the Crystal Barrel data
the value is 12.2~$\pm$~5.4.
A coupled channel fit of the $\pi \pi$, $K \overline K$, $4\pi$, $\eta \eta$
and $\eta \eta^\prime$ final states is in progress and will
be reported in a future publication.
\par
In summary,
the $dP_T$, $\phi$ and $|t|$ distributions for the $f_0(1370)$,
$f_0(1500)$, $f_0(2000)$ and $f_2(1950)$ have been presented.
For the $J$~=~0 states the $f_0(1370)$ and $f_0(2000)$
have similar $dP_T$ and $\phi$ dependences. These are different
to the $dP_T$ and $\phi$ dependences of the $f_0(980)$, $f_0(1500)$
and $f_0(1710)$. For the $J$~=~2 states the
$f_2(1950)$ has different dependences to the
$f_2(1270)$ and $f_2^\prime(1520)$.
This shows that the $dP_T$ and $\phi$ dependences are not just $J$ phenomena.
\begin{center}
{\bf Acknowledgements}
\end{center}
\par
This work is supported, in part, by grants from
the British Particle Physics and Astronomy Research Council,
the British Royal Society,
the Ministry of Education, Science, Sports and Culture of Japan
(grants no. 07044098 and 1004100), the French Programme International
de Cooperation Scientifique (grant no. 576)
and
the Russian Foundation for Basic Research
(grants 96-15-96633 and 98-02-22032).

\newpage
\begin{table}[h]
\caption{Production of the resonances as a function of $dP_T$
expressed as a percentage of their total contribution and the
ratio (R) of events produced at $dP_T$~$\leq$~0.2~GeV to the events
produced at $dP_T$~$\geq$~0.5~GeV.}
\label{ta:dpt}
\vspace{1in}
\begin{center}
\begin{tabular}{|c|c|c|c|c|} \hline
 & & & & \\
 &$dP_T$$\leq$0.2 GeV & 0.2$\leq$$dP_T$$\leq$0.5 GeV &$dP_T$$\geq$0.5 GeV &
$R=\frac{dP_T \leq 0.2 GeV}{dP_T\geq 0.5 GeV}$\\
 & & & & \\ \hline
 & & & & \\
$f_0(1370)$  &11.0 $\pm$ 2.0 & 32.9 $\pm$ 3.0  &56.1 $\pm$ 4.9  &
0.19~$\pm$~0.04\\
 & & & & \\ \hline
 & & & & \\
$f_0(1500)$  &23.8 $\pm$ 2.5  & 47.3 $\pm$ 4.5  &28.8 $\pm$ 2.9   &
0.83~$\pm$~0.12 \\
 & & & & \\ \hline
 & & & & \\
$f_0(2000)$  &11.9 $\pm$ 1.3  & 37.7 $\pm$ 3.2 &50.2 $\pm$ 4.1 &
0.23~$\pm$~0.03\\
 & & & & \\ \hline
 & & & & \\
$f_2(1950)$  &27.4 $\pm$ 2.4  & 45.5 $\pm$ 5.1 &27.1 $\pm$ 2.4 &
1.01~$\pm$~0.12\\
 & & & & \\ \hline
\end{tabular}
\end{center}
\end{table}
\begin{table}[h]
\caption{The slope parameter $b$ from a single exponential fit to the $|t|$
distributions.}
\label{ta:b}
\vspace{0.3in}
\begin{center}
\begin{tabular}{|c|c|c|c|c|} \hline
  &&&& \\
 & $f_0(1370)$ & $f_0(1500)$ & $f_0(2000)$ & $f_2(1950)$\\
   &&&&\\ \hline
  &&&& \\
 b/GeV$^{-2}$ & $5.8 \pm 0.5$ & $5.1 \pm 0.4$ & $5.6 \pm 0.4$ & $5.9 \pm 0.4$
\\
   &&&& \\ \hline
\end{tabular}
\end{center}
\end{table}
\clearpage
{ \large \bf Figures \rm}
\begin{figure}[h]
\caption{
The $J^{PC}$~=$0^{++}$~$\rho \rho$ wave from the \pic channel
as a function of $\phi$.
a) $\phi$~$<$ 45 degrees,
b) 45 $<$~$\phi$~$<$ 90 degrees,
c) 90 $<$~$\phi$~$<$ 135 degrees and
d) 135 $<$~$\phi$~$<$ 180 degrees.
The superimposed curves are the resonance contributions coming from
the fits described in the text.}
\label{fi:1}
\end{figure}
\begin{figure}[h]
\caption{The $\phi$ and
four momentum transfer squared ($|t|$)
distributions for
a), b) the $f_0(1370)$,
c), d) the $f_0(1500)$,
e), f) the $f_0(2000)$ and
g), h) the $f_2(1950)$.
}
\label{fi:2}
\end{figure}
\begin{figure}[h]
\caption{
a) and b) The $\phi \phi$  and
c) the $K^* \overline K^*$ mass spectra with fits described in the text.
}
\label{fi:3}
\end{figure}
\begin{center}
\epsfig{figure=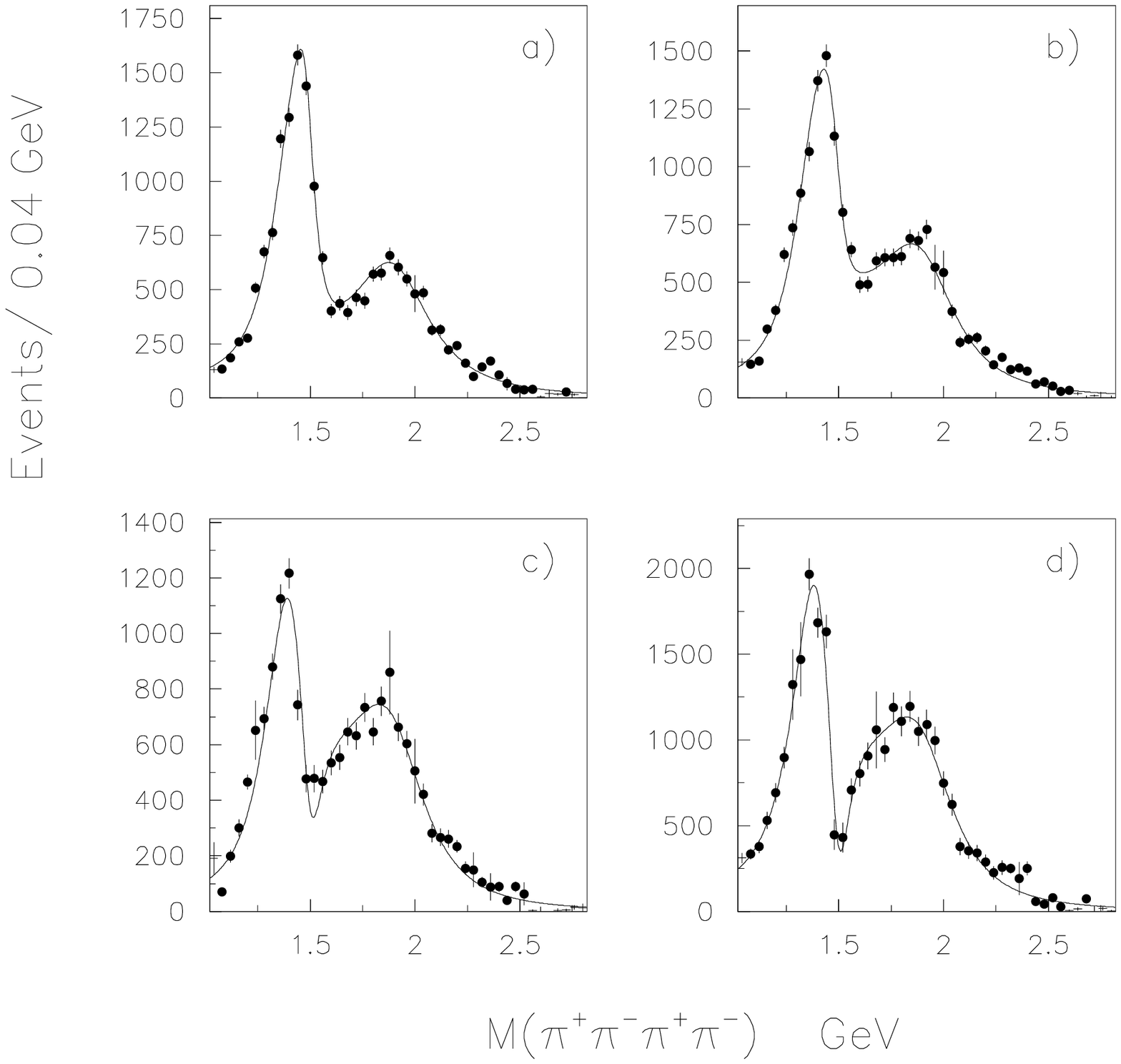,height=22cm,width=17cm}
\end{center}
\begin{center} {Figure 1} \end{center}
\newpage
\begin{center}
\epsfig{figure=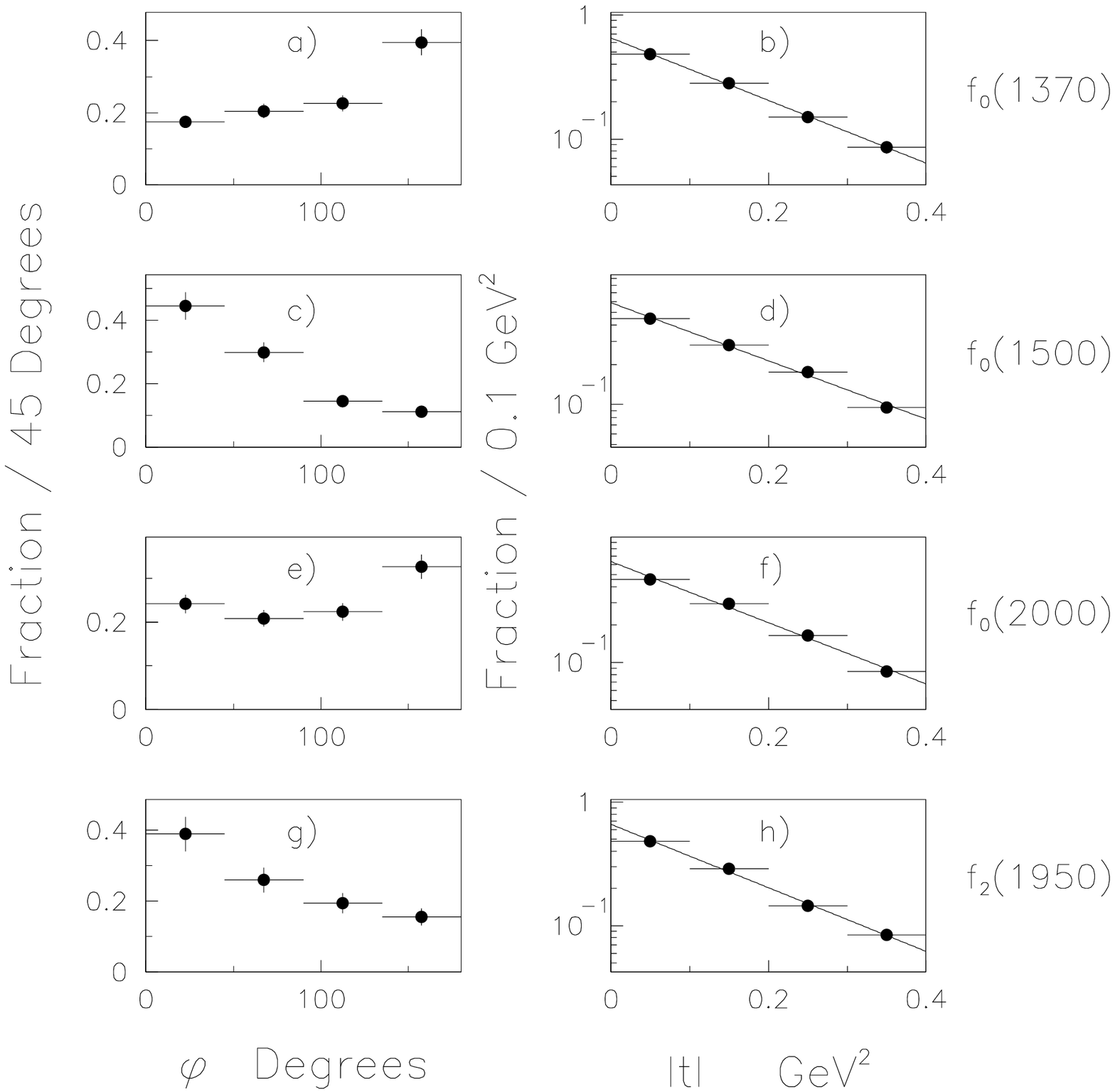,height=22cm,width=17cm}
\end{center}
\begin{center} {Figure 2} \end{center}
\newpage
\begin{center}
\epsfig{figure=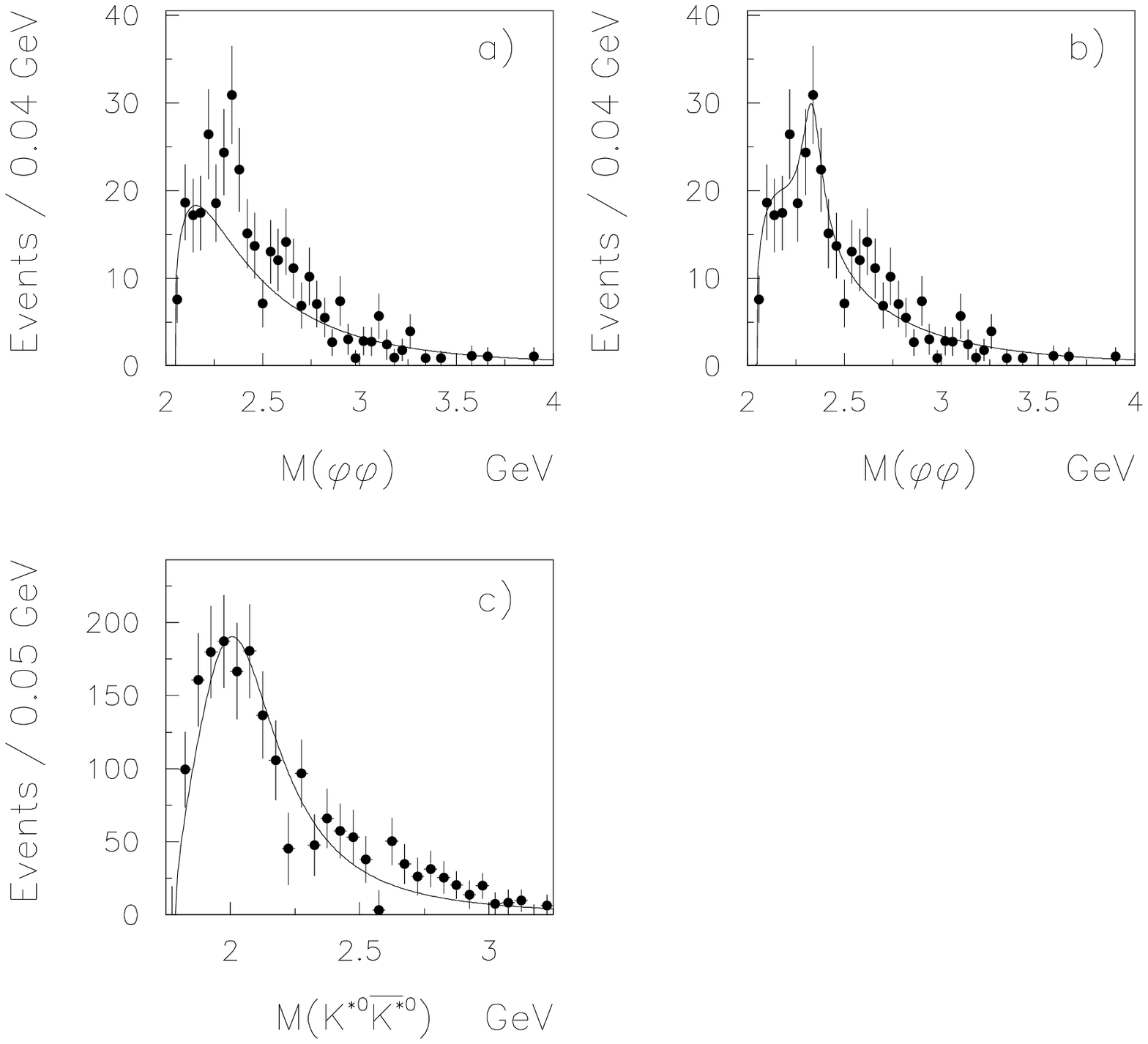,height=22cm,width=17cm}
\end{center}
\begin{center} {Figure 3} \end{center}
\end{document}